# Modelling of Current Percolation Channels in Emerging Resistive Switching Elements

Marcus Wu Shihong, Themistoklis Prodromakis, *Member, IEEE*, Iulia Salaoru, *Member, IEEE*, Christofer Toumazou, *Fellow,IEEE*

*Abstract*— **Metallic oxides encased within Metal-Insulator-Metal (MIM) structures can demonstrate both unipolar and bipolar switching mechanisms, rendering them the capability to exhibit a multitude of resistive states and ultimately function as memory elements. Identifying the vital physical mechanisms behind resistive switching can enable these devices to be utilized more efficiently, reliably and in the long-term. In this paper, we present a new approach for analysing resistive switching by modelling the active core of two terminal devices as 2D and 3D grid circuit breaker networks. This model is employed to demonstrate that substantial resistive switching can only be supported by the formation of continuous current percolation channels, while multi-state capacity is ascribed to the establishment and annihilation of multiple channels.**

## I. INTRODUCTION

RESISTIVE switching (RS) is the reversible change in the resistance of a material that has been observed in many perovskite oxides and binary metal oxides encased within Metal-Insulator-Metal (MIM) structures [1]. Depending on the polarity of the set and reset potentials required to toggle between distinct states, RS can be classified as unipolar (URS) or bipolar (BRS) [2]. Such devices typically exhibit a pinched hysteresis loop in the current-voltage domain, which has been shown to be the signifying property of a class of devices nowadays known as memristors [3], irrespectively of the constituent materials and nature of switching mechanisms [4], [5].

Particularly in recent years, RS finds applications in reconfigurable architectures [6] and neuromorphic computing [7], leveraging devices with miniscule dimensions that possess capacity for multi-level state programming [8]. Therefore, deciphering the physical mechanisms of RS is imperative to heralding memristors as emerging devices; particularly in devices that extend Moore's scaling trend [9] beyond its current physical limits.

This work was partially supported by Mr. Wilfred J. Corrigan, the CHIST-ERA ERA-Net, EPSRC EP/J00801X/1, and the Defense Science and Technology Agency (DSTA) of Singapore.

Marcus Wu Shihong, T. Prodromakis, I. Salaoru and C. Toumazou are with the Centre for Bio-Inspired Technology, Department of Electrical and Electronic Engineering, Imperial College London, South Kensington Campus, London SW7 2AZ, UK (e-mail: marcus.wu-shihong08@imperial.ac.uk).

The functional properties of such devices have been ascribed to Joule heating [10], electrochemical migration of oxygen ions [11] and vacancies [12], the lowering of Schottky barrier heights by trapped charge carriers at interfacial states [13], the phase-change [14], [15] and the formation and rupture of conductive filaments [16-18] in a device's active core. Likewise, a plurality of models has been proposed for each distinct scenario. While there is no single encompassing model that can explain all observations, newer models have demonstrated increasing accuracy and relevancy in their results [19].

We prove here that substantial RS is only viable through the formation and annihilation of continuous conductive percolation channels between the top (TE) and bottom (BE) electrode of a device. This study is facilitated via empirical circuit breaker network models [20] and our results are compared against existing model systems such as Hewlett Packard's (HP) memristor [21]. An overview of the grid circuit network model along with configuration descriptions are given in Section II. In Section III, we study 2D grid networks and demonstrate the effect on RS of discrete conducting branches and continuous current percolation channels in a device's core. Finally, we extend this study by employing 3D grids and presenting the relevancy of these results.

## II. GRID CIRCUIT NETWORK MODEL

The term 'memristor' is used interchangeably with the term 'memory element', as well as in reference to the entity modelled by the grid circuit network. Memristors typically comprise a block of an insulating metal-oxide layer sandwiched between two metal layers that serve as the TE and BE. The metal-oxide layer can be represented by a grid circuit network as shown in Fig. 1. Assuming the layer is stoichiometric in nature, this grid is evenly distributed within its dimensions. Each horizontal and vertical branch in the network can be made up of a single circuit element or a combination of other elements that describe the unit area impedance of the material. The branches are interconnected at the nodes to form a complex mesh network with the lines in solid black denoting the TE and BE.

Here we specifically employ grid circuit breaker network models comprising static resistors to simulate the resistive behaviour of a metal-oxide cross-section layer. High resistances typically represent bulk regions of insulating material while low resistances represent localized areas in which conductive percolation channels exist. While the metal-oxide layer in these devices is generally insulating, percolation channels demonstrate metallic conductivity at room temperature [22] and can vastly alter the overall resistance of the device.

Every static resistor has a pre-chosen resistance value. For a block of metal-oxide in its high resistive state (HRS), every resistor will be of the same resistance. Supplying a sufficient potential alters the resistive state of the device, and this local modification of the active layer is represented by the alteration of some branch resistances to higher or lower values. Individual branches with reduced resistances are known as percolation branches and when they are connected continuously from TE to BE, they establish a percolation channel. This channel is essentially a high conductance path that greatly reduces the Thevenin resistance of the grid network. Therefore, we infer that significant RS is only possible if a percolation channel exists, no matter which mechanism is responsible for its establishment.

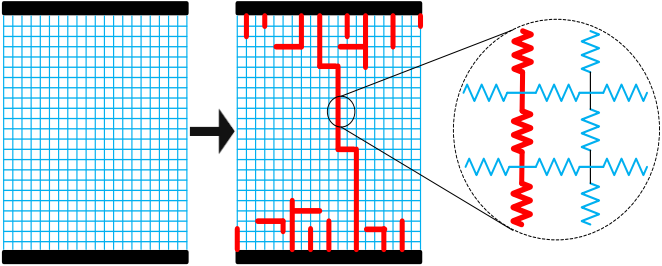

**Figure 1**: *Representing a block of memory material as a grid circuit network. The material is assumed to be insulating and initially in HRS. Applying a potential at the TE induces the formation of low resistance percolation branches and channels (in red) within a grid network of high resistance branches (in blue).*

Distinct resistive states of any two-terminal memory element can be supported by varying the number of percolation channels within the grid network. In conducting simulations, resistance values had to be chosen for the high resistance branches and percolation branches. The high resistance value used to represent the branches in HRS may vary as the overall resistance of the device is dependent on its fabricated dimensions and electrical permittivity. This value may also change as the active layer transits from one resistive state to another, or to accommodate a material's degradation due to exposure in high electric field densities.

On the other hand, resistive elements representing the conductive percolating branch have been fixed. Throughout this paper, we emulate the formation of each percolation branch as a closing of a single quantum point contact (QPC) switch [23]. As the resistance of memory devices has been observed to change in quantized steps upon the application of voltage pulses [24], the transition of a single high resistance branch to a high conductance percolation branch and vice versa could serve as a plausible explanation for the change from one quantized resistive state to another. This phenomenon has also been observed in many gap-type devices [25-27], when two electrodes are separated by a nm-scale gap.

While the exact constituents of these filaments are yet to be determined [28], their presence can generally be attributed to the formation and dissolution of atomic bridges. Assuming a nanometre-scale device, a conductive filament can be approximated by a single metal atom bridge, which is typically formed by the vertical stacking of many atoms in contact from TE to BE [29]. Equivalently, this can be represented by a number of serially connected resistors, which denotes the percolation channel in our grid circuit network model. In this case, each percolation branch is set as a quantum resistance $R_0$, which is commonly known as Landauer conductance $G_0$ [30]. $R_0$ is expressed as a function of Planck's constant $h$ and electron charge $e$:

$$R_0 = \frac{h}{2e^2} = 12.9k \text{ or } G_0 = \frac{1}{R_0} = 77.4\ \mu S \quad (1)$$

The overall resistance of an atomic bridge composed of $n$ atoms is equal to the series sum of quantum resistances:

$$R_{bridge} = \sum_{i=1}^{n} R_0 = 12.9n\ k\Omega \quad (2)$$

Clearly, this approach is rather versatile since distinct RS mechanisms can be described via "local bridging" that facilitates current percolation, no matter what the underlying switching mechanism is [31].

## III. APPLYING GRID CIRCUIT NETWORK TO SIMULATE HP'S MEMRISTOR

In 2008, HP demonstrated the aptitude of $TiO_2$ in RS [32] and correlated this response with the theoretical background on memristors [3] presented more than 3 decades ago. The multitude of resistive states exhibited by a voltage-driven $TiO_{2-x}/TiO_2$ bi-layer of thickness $D$ was ascribed to the drift of oxygen vacancies that causes modulation of the thickness $w$ of the oxygen-deficient $TiO_{2-x}$ layer. As the $TiO_{2-x}$ layer is more conductive as compared to the adjacent $TiO_2$ layer, the device was emulated as two variable resistors in series [32], each one accounting for the distinct conductivities of the $TiO_{2-x}$ and $TiO_2$ bi-layers. We first investigated whether the change in thickness of the conductive layer is sufficient in producing substantial RS. We modelled the $TiO_{2-x}/TiO_2$ layers of the device with a 10x10 grid circuit network, where the layers can be defined by segmenting the grid horizontally into two regions, comprising high and low resistive branches. Dimensions are defined by the number of nodes and a DC voltage was applied to the networks in a simulations.

In the pristine state, the $TiO_2$ and $TiO_{2-x}$ layers are initially distributed evenly ($w=D/2$). The application of a positive (negative) bias voltage will increase (decrease) the width $w$ of the conductive $TiO_{2-x}$ layer. Considering every unit branch in the resistive $TiO_2$ and conductive $TiO_{2-x}$ layer as 60MΩ and 12.9kΩ respectively, the initial resistance of the device in Fig. 2(a) is 36MΩ. Fig 2(b) shows the corresponding "ON" state model, where 90% of the branches have become more conductive while Fig. 2(c) shows a low-resistive state (LRS) model, where only 20% of the branches are conductive.

We assumed that the "OFF" state is where every branch in the grid is highly resistive, giving an overall resistance of 66MΩ. The calculated overall resistance of the device as described by the networks presented in Figs. 2(b) and 2(c) are correspondingly 6MΩ and 48MΩ and the resulting $R_{OFF}/R_{ON}$ ratio is 11. Although we have substantially varied the width $w$ ($0.2D \leq w \leq 0.9D$) of the layer in which the mobile ions have congregated, we were only able to observe an order of magnitude change in the device's static resistance. We then modified our model as shown in Fig. 2(d) to account for one continuous percolation channel. In this case, the overall resistance of the network has decreased to 43kΩ, giving a much higher $R_{OFF}/R_{ON}$ ratio of 694. The addition of this branch is equivalent to the closing of a QPC switch and together with the entire conductive layer, it forms the conductive path required to produce significant RS.

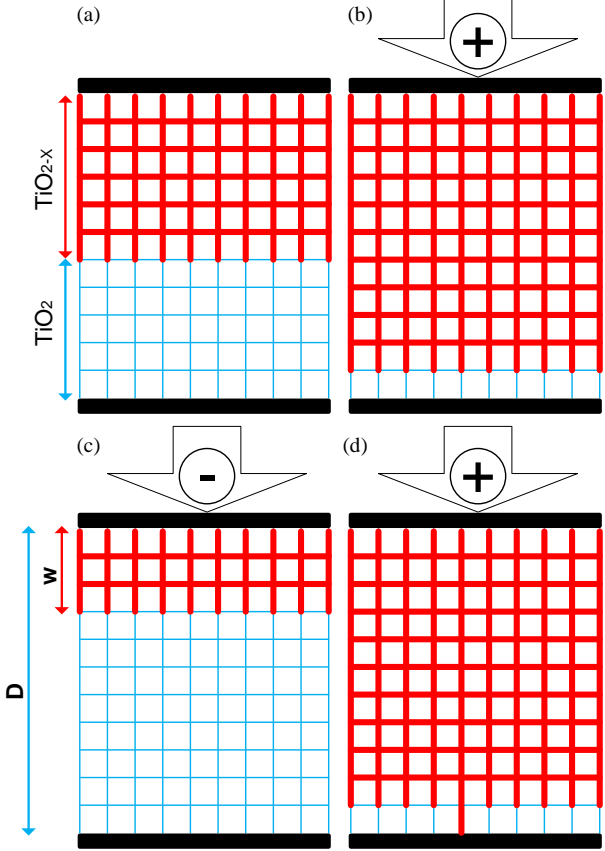

*Figure 2: Applying the Grid Circuit Network to HP's physical model of a $TiO_{2-x}/TiO_2$ bi-layer. (a) $TiO_{2-x}$ and $TiO_2$ layers were initially distributed evenly, denoted by equal number of quantum resistance branches (in red) and high resistance branches (in blue). (b) Applying a positive bias increases the width of the conductive $TiO_{2-x}$ layer. (c) Applying a negative bias decreases the width of the conductive $TiO_{2-x}$ layer. (d) Formation of a continuous percolation channel from TE to BE through the addition of one conductive branch.*

To determine the isolated effect of the percolation channel, we simulated different cases where RS is dependent upon the following: (1) thickness of the initial conductive layer, (2) the type of conductive path taken, (3) the effect of randomly occurring percolation branches and (4) the number of percolation channels. We first considered the case where the device is initially in the "OFF" state, depicted by Fig. 2(c). Assuming that the positive bias voltage produces a straight, continuous percolation channel as shown in Fig. 3(a), the overall resistance will decrease from 48MΩ to 124kΩ, providing a $R_{OFF}/R_{ON}$ ratio of 387.

To determine the impact of the conductive layers on the overall resistance, these were removed and replaced by a lone straight percolation channel through the insulating layer, as shown in Fig. 3(b). With every branch in the percolation channel rated as a Landauer conductance, the overall resistance of this network decreased from 66MΩ to 142kΩ, resulting in a $R_{OFF}/R_{ON}$ ratio of 465. Similarly to the scenario depicted in Fig. 2(d), we can conclude that the presence of conductive layers in the device's core indeed influences the magnitude of RS. Nonetheless, the effect of forming and annihilating continuous percolation channels is much more substantial, facilitating $R_{OFF}/R_{ON}$ ratios of several orders of magnitude. Besides, we have determined that the path taken by the percolation channels also impacts the overall resistance of the device. A winding, non-uniform percolation channel as shown in Fig. 3(c) results in an overall resistance of 270kΩ, yielding a ratio of 245. Similarly, the random occurrence of individual percolation branches within the grid, as shown in Fig. 3(d), has minimal impact so long as the percolation channel remained present. In this case, the resistance decreased marginally to 266kΩ and gave a slightly higher ratio of 248.

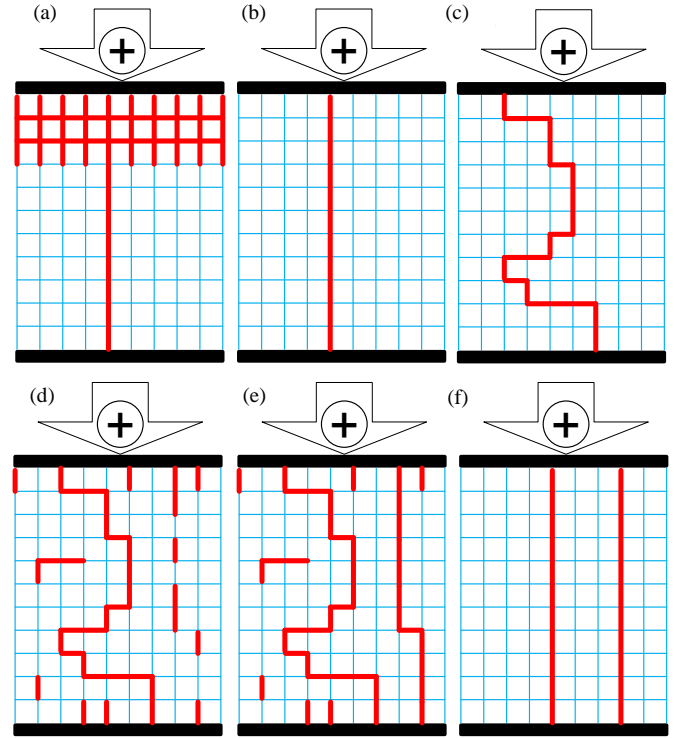

*Figure 3: Simulated formation of percolation channels through the grid circuit network. (a) Formation of a single straight percolation channel to switch the device from the "OFF" state shown in Fig. 2(c) to an "ON" state. (b) Formation of a single percolation channel without any conductive layers at the TE. (c) Formation of a winding, non-uniform percolation channel. (d) Winding, non-uniform percolation channel with randomly occurring individual percolation branches. (e) Individual percolation branches shown in Fig. 4(d) are connected to form a second percolation channel. (f) Formation of two straight percolation channels without a conductive layer and other individual percolation branches.*

Next, we investigated the influence of percolation channel positioning within the grid by deploying three straight paths that become increasingly spaced out. In Fig. 4(a), the three channels are adjacent to one another and the resistance was evaluated to be 47.3 k$\Omega$, giving a ratio of 1395. The second and third channels were then spaced out slightly further as illustrated in Figs. 4(b) and 4(c). However, a negligible change in the overall resistance and corresponding $R_{OFF}/R_{ON}$ ratios was observed. This demonstrates that the physical location where the filamentary formation takes place, although important for facilitating RS [34], has no significant implication to the extent of switching.

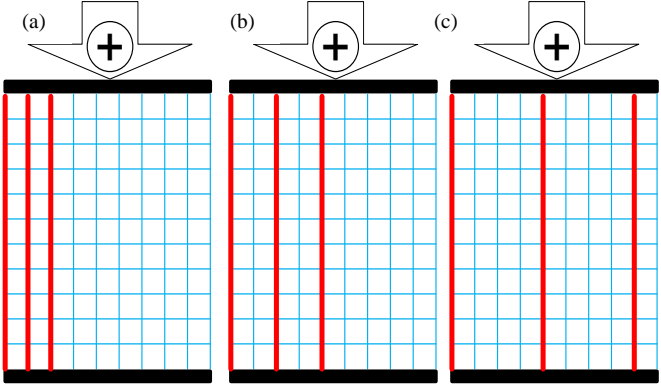

*Figure 4*: *Simulating effects of positioning three percolation channels in different ways, which was found to have no effect on the overall resistance of the device (a) Adjacent to one another. (b) Spaced out by 1 column. (c) Spaced out by 3 columns.*

Concurrently, the ratios for Figs. 4(a)-(c) are approximately three times as much as that of Fig. 3(b). Hence, we empirically verified that the $R_{OFF}/R_{ON}$ ratio increases with the number of percolation channels. Considering the ratio in the presence of one straight channel as $(R_{OFF}/R_{ON})_{(1)}$ and the number of channels as $n$, an empirical formula for the $R_{OFF}/R_{ON}$ ratio of the device is $n(R_{OFF}/R_{ON})_{(1)}$. We also simulated cases with 4 and 5 percolation channels and the obtained ratios were correspondingly 1860 and 2324 respectively. This trend is illustrated in Fig. 5 and the linear relationship obtained verifies our empirical formula. Clearly, the ramification of this observation is that multistate capacity of memristors is facilitated by the formation and annihilation of multiple current percolation channels in a device's core.

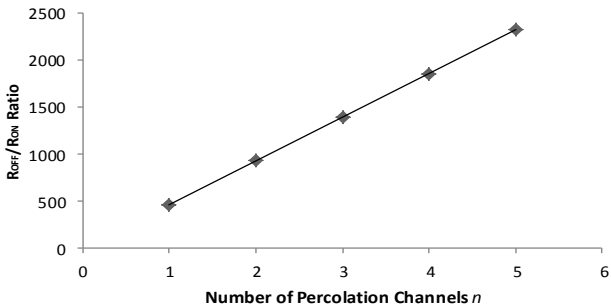

*Figure 5*: *Effect of number of current percolation channels on $R_{OFF}/R_{ON}$ ratio of a model device.*

## IV. MODELLING 3D GRID NETWORKS

We have extended this study to account for 3D grid circuit networks (10x10x5), where every unit branch has a resistance of 60M$\Omega$. The initial resistance of the layer in its "OFF" state is 12M$\Omega$. A total of 6 percolation channels were placed at random locations one after another successively within the network as shown in Fig. 6 and discrete percolation branches were also added.

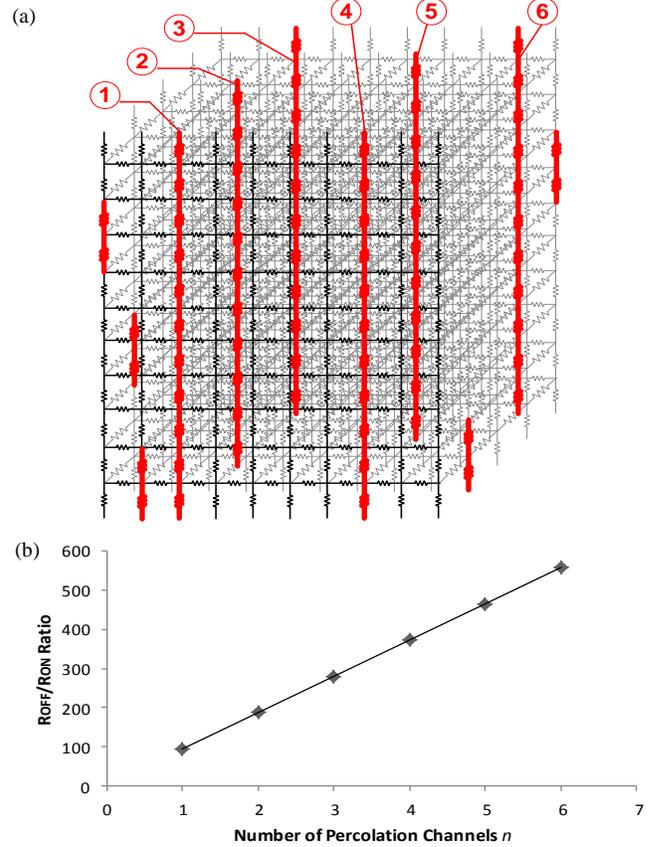

*Figure 6*: *(a) Sequential addition of individual percolation channels (numbered 1 to 6) and conductive branches at random locations within a 10x10x5 grid circuit breaker network. Linear relationship between number of current percolation channels and $R_{OFF}/R_{ON}$ ratio is verified in (b).*

The overall resistance decreased sequentially from 12M$\Omega$ to 128k$\Omega$, to 64k$\Omega$, to 43k$\Omega$, to 32k$\Omega$, to 26k$\Omega$ and finally to 21k$\Omega$. The $R_{OFF}/R_{ON}$ ratio was also calculated to have increased from 94, to 187, to 280, to 373, to 466 and finally to 558. The $R_{OFF}/R_{ON}$ ratios were plotted against the number of percolation channels and the empirical formula was yet again verified by the linear relationship previously obtained. As such, we conclude that the network can exist in 6 different resistive states if a total of 6 percolation channels are formed in succession. Therefore, a device layer can modelled in both two and three dimensions and varying the number of percolation channels in either network will still allow a modelled device to exhibit substantial RS and multi-state capacity. Modelling with 3D grid circuit networks also provides a more realistic modelling approach as the grids can be optimized to map the dimensions of practical implementations.

## V. CONCLUSION

By modelling the active core of a two-terminal MIM-based device with 2D and 3D grid circuit breaker networks, we demonstrated that RS and the multitude of resistive states exhibited by memory devices can mainly be explained by the formation and annihilation of conductive percolation channels. Apart from their presence, the number of channels, the paths taken by these channels, the presence of conductive layers and individual percolation channels also influence the resistive state of the device. We also proved that altering the thickness of the conductive layer alone cannot account for the substantial RS involved. Therefore, we believe that substantial RS in perovskite and binary metal oxides can only occur in the presence of conductive percolation channels no matter what the underlying mechanism or the material in support of this mechanism is. Finally, it was demonstrated that the multistate capacity of emerging memory devices is solely ascribed to the establishment and annihilation of multiple current percolation channels.